%% file: ms.tex
\documentclass[extra]{gji}
\pdfoutput=1
\usepackage{timet}
\usepackage{lineno}
\usepackage{graphicx}
\usepackage{amsmath,amsfonts,amssymb}
\usepackage{breqn}
\usepackage[normalem]{ulem}
\usepackage[dvipsinames]{xcolor}
\usepackage{longtable}

\usepackage[%
           colorlinks=True,%
           urlcolor=blue,%
           linkcolor=blue,%
           citecolor= blue,%
           pdfauthor={T. Schwaiger},%
           pdftitle={Force balance in numerical geodynamo simulations: a systematic study},]{hyperref}
\usepackage{booktabs}
\usepackage{hyperref}

\definecolor{darkgreen}{rgb}{0,0.6,0}

\let\oldalign\align
\let\oldendalign\endalign
\renewenvironment{align}
  {\linenomathNonumbers\oldalign}
  {\oldendalign\endlinenomath}

\title[Force balance in numerical geodynamo simulations: a systematic study]
  {Force balance in numerical geodynamo simulations: a systematic study}
\author[T. Schwaiger et al.]
  {T. Schwaiger$^1$, T. Gastine$^1$ and J. Aubert$^1$ \\
  $^1$ Institut de Physique du Globe de Paris, Sorbonne Paris Cit\'e, Universit\'e Paris-Diderot, CNRS, 1 rue Jussieu, F-75005 Paris, France.
  }

\date{Received \today; in original form \today}
\pagerange{\pageref{firstpage}--\pageref{lastpage}}

\begin{document}

\label{firstpage}

\maketitle

\begin{summary}
Dynamo action in the Earth's outer core is expected to be controlled by a balance between pressure, Coriolis, buoyancy and Lorentz forces, with marginal contributions from inertia and viscous forces. Current numerical simulations of the geodynamo, however, operate at much larger inertia and viscosity because of computational limitations. This casts some doubt on the physical relevance of these models.

Our work aims at finding dynamo models in a moderate computational regime which reproduce the leading-order force balance of the Earth. By performing a systematic parameter space survey with Ekman numbers in the range $10^{-6} \leq E \leq 10^{-4}$, we study the variations of the force balance when changing the forcing (Rayleigh number, $Ra$) and the ratio between viscous and magnetic diffusivities (magnetic Prandtl number, $Pm$). For dipole-dominated dynamos, we observe that the force balance is structurally robust throughout the investigated parameter space, exhibiting a quasi-geostrophic (QG) balance (balance between Coriolis and pressure forces) at zeroth order, followed by a first-order MAC balance between the ageostrophic Coriolis, buoyancy and Lorentz forces. At second order this balance is disturbed by contributions from inertia and viscous forces. Dynamos with a different sequence of the forces, where inertia and/or viscosity replace the Lorentz force in the first-order force balance, can only be found close to the onset of dynamo action and in the multipolar regime. To assess the agreement of the model force balance with that expected in the Earth's core, we introduce a parameter quantifying the distance between the first- and second-order forces. Analysis of this parameter shows that the strongest-field dynamos can be obtained close to the onset of convection ($Ra$ close to critical) and in situations of reduced magnetic diffusivity (high $Pm$). Decreasing the Ekman number gradually expands this regime towards higher supercriticalities and lower values of $Pm$.

Our study illustrates that most classical numerical dynamos are controlled by a QG-MAC balance, while cases where viscosity and inertia play a dominant role are the exception rather than the norm.
\end{summary}


\section{Introduction}
The Earth's magnetic field is believed to be generated by dynamo action in the liquid outer core. The flow dynamics driving this process are expected to be controlled by a balance between pressure, Coriolis, buoyancy and Lorentz forces, with marginal contributions from inertia and viscous forces \citep[e.g.,][]{roberts_and_king_2013}. However, the exact structure of the leading-order force balance is still debated \citep[e.g.,][]{dormy_2016, aubert_etal_2017, aurnou_and_king_2017}.

Historically, theoretical considerations largely based on asymptotic studies of magneto-convection resulted in the distinction between weak- and strong-field regimes of dynamo action \citep[e.g.,][]{hollerbach_1996}. In a system dominated by rapid rotation, as is the case for the Earth's core, fluid motions tend to be invariant in the direction of the rotation axis, and fulfill the so-called Proudman-Taylor constraint. In the weak-field regime, this rotational constraint is broken by the viscous force or inertia, while the Lorentz force is substantially weaker, leading to small-scale convection. Increasing the vigour of convection increases the magnetic field strength and as a result, the Lorentz force could eventually break the rotational constraint, leading to larger convective scales.
This induces a catastrophic runaway growth of the magnetic field until convection occurs on the scale of the system size. At this point the magnetic field equilibrates at Elsasser number $\Lambda \sim \mathcal{O}(1)$, where $\Lambda$ measures the relative amplitudes of the Lorentz and Coriolis forces:
\begin{align}
\Lambda = \frac{\vert F_{\mathrm{Lorentz}} \vert }{\vert F_{\mathrm{Coriolis}} \vert} = \frac{\vert \mathbf{J} \times \mathbf{B} \vert}{\vert 2 \rho \mathbf{\Omega} \times \mathbf{u} \vert} \sim \frac{J B}{\rho \Omega U},
\end{align}
with $J$ representing the current density, $B$ the magnetic field strength, $\rho$ the fluid density, $\Omega$ the rotation rate and $U$ the flow velocity.
The resulting regime is referred to as strong-field regime due to the Lorentz force now being much stronger than the viscous forces and inertia. Since a magnetic field with $\Lambda \sim \mathcal{O}(1)$ facilitates convection \citep{malkus_1959}, it has been suggested that the core flow dynamics are in a magnetostrophic (MS) state, where pressure, Coriolis and Lorentz forces balance each other at zeroth order \citep{wu_and_roberts_2013}. To assess based on geomagnetic observations whether the geodynamo operates in a magnetostrophic regime, the Elsasser number has traditionally been estimated using the following definition:
\begin{align}
\Lambda_{t} = \frac{B^{2}}{\rho \mu \eta \Omega},
\end{align}
where $\mu$ represents the magnetic permeability and $\eta$ the magnetic diffusivity. Inserting characteristic values of the Earth \citep[e.g.,][]{christensen_and_aubert_2006} yields ${\Lambda_{t}\sim\mathcal{O}(1)}$, which has often been used to argue for Lorentz and Coriolis forces being of the same order of magnitude in the outer core. The definition of $\Lambda_{t}$, however, does not include length and velocity scales and therefore may provide an inaccurate measure of the relative amplitudes of the forces \citep{soderlund_etal_2012, soderlund_etal_2015, calkins_2018}. A more rigorous estimate of this force ratio can be obtained using a dynamic Elsasser number, defined as \citep{soderlund_etal_2012}:
\begin{align}
\Lambda_{d} = \frac{B^{2}}{\rho \mu \Omega U D},
\end{align}
where $D$ represents the thickness of the outer core. Employing characteristic values to this definition yields ${\Lambda_{d}\sim\mathcal{O}(10^{-2})}$ for the Earth, indicating that the Lorentz force is two orders of magnitude smaller than the Coriolis force. This suggests that core flow dynamics could be controlled by a geostrophic balance between pressure and the Coriolis force at leading order, which would result in quasi-geostrophic (QG) instead of magnetostrophic convection dynamics \citep{soderlund_etal_2012, calkins_2018}. This is in agreement with recent studies of the core flow based on the inversion of geomagnetic secular variation data, which suggest that on global scales QG flows appear to describe the observations best \citep[e.g.,][]{gillet_etal_2012}. Theoretical grounding for how such large-scale quasi-geostrophy could be possible has recently been provided by \citet{aurnou_and_king_2017}, who suggested based on scaling analysis of the Elsasser number, that magnetostrophic flow dynamics may be deferred to smaller scales, inaccessible to geomagnetic observations. \citet{aurnou_and_king_2017} therefore argue for a length scale dependent combination of zeroth-order quasi-geostrophy and magnetostrophy.

In addition to observations and theoretical considerations, global numerical dynamo simulations represent an important tool for our understanding of the dynamo mechanism. Although computational resources have increased significantly since the first successful dynamo simulations computed more than 20 years ago \citep{glatzmaier_and_roberts_1995}, current numerical models still operate at parameters far from the expected conditions of the Earth's core. Despite this limitation, numerical dynamos have proven to be very successful in reproducing numerous features of the geomagnetic field \citep[e.g.,][]{Christensen_etal_2010}. However, it remains uncertain whether these results are obtained for the right physical reasons. As a consequence many studies have been performed, trying to answer this question.
\citet{soderlund_etal_2012} found that convection in many dynamo simulations does not occur on the system scale, but rather on a scale similar to that of rotating convection. As a result, they argued for a subdominant role of the magnetic field in those numerical models, which was attributed to a sizeable contribution of viscosity \citep[e.g.,][]{king_and_buffett_2013, oruba_and_dormy_2014}. Some authors even suggested that the majority of dynamos found to date belong to the viscous weak-field regime \citep[e.g.,][]{dormy_2016, dormy_etal_2018}. However, the discrepancy between dynamo solutions and non-magnetic convection was shown to become more obvious when the viscosity in the models is lowered \citep[e.g.,][]{sakuraba_and_roberts_2009, yadav_etal_2016}.
By explicitly computing the magnitude of all forces \citep[e.g.,][]{wicht_and_christensen_2010,soderlund_etal_2012, soderlund_etal_2015} and their level of cancellation \citep[e.g.,][]{yadav_etal_2016}, numerical dynamos were found to be quasi-geostrophic at zeroth order, with buoyancy and Lorentz forces balancing the ageostrophic Coriolis force, i.e. the part of the Coriolis force which is not balanced by pressure. More recently, \citet{aubert_etal_2017} introduced a length scale dependent approach for a more refined analysis of the force balance. High resolution dynamos \citep[e.g.,][]{yadav_etal_2016, schaeffer_etal_2017, sheyko_etal_2018} at advanced parameter regimes support this QG-MAC (Quasi-Geostrophic Magneto-Archimedean-Coriolis) balance, which suggests that it could go all the way to the core \citep{aubert_etal_2017}.

While the majority of dynamo models to date therefore seem to support a QG-MAC balance, some studies also report dynamos that could be controlled by a magnetostrophic balance at zeroth order \citep[e.g.,][]{dormy_2016, dormy_etal_2018}. \citet{dormy_2016} and \citet{dormy_etal_2018} were further able to observe close to the onset of dynamo action a catastrophic runaway growth of the magnetic field from viscously dominated weak-field dynamos to strong-field dynamos in their models, similar to that predicted by asymptotic studies of rotating magneto-convection. These conflicting interpretations illustrate that not only the presumed force balance in the Earth's core but also the force balance obtained in current numerical models are still highly-debated topics. 

By performing a systematic survey of the numerically accessible parameter space, we attempt to enable a better understanding of force balances. To this end, we will make use of the scale dependent force balance representations introduced in \citet{aubert_etal_2017}. Additionally, we will introduce new tools to directly relate the physical scale at which the dynamo is organised locally to the governing force balance.
Throughout this work, we adopt the following naming conventions. Some authors use the term `magnetostrophic' (strictly) to describe a zeroth-order MS balance \citep[e.g.,][]{roberts_1978, dormy_2016, dormy_etal_2018}. Other authors consider `magnetostrophic' as the QG-MAC balance \citep[e.g.,][]{yadav_etal_2016, aubert_etal_2017}. Here, `magnetostrophic' will only refer to the zeroth-order MS balance to avoid possible confusion between the two types of force balances. Likewise, some authors refer to dynamos as being in the strong-field regime only in the presence of an $\Lambda\sim\mathcal{O}(1)$ magnetic field, i.e. dynamos controlled by a zeroth-order MS balance \citep[e.g.,][]{roberts_1978, dormy_2016, dormy_etal_2018}. Many other authors consider strong field simply as the magnetic energy being much larger than the kinetic energy \citep[e.g.,][]{schaeffer_etal_2017, aubert_etal_2017}. This is the definition that we will retain here.

Section 2 presents the numerical models and methods. The results are presented in section 3, followed by a discussion in section 4.

\section{Numerical model}
We consider a spherical shell of thickness $D=r_{o}-r_{i}$ and radius ratio $r_{i}/r_{o} = 0.35$, where $r_i$ and $r_{o}$ are the inner and outer radii. The shell rotates with angular frequency $\Omega$ about the axis $\mathbf{e}_{z}$. The inclosed fluid of density $\rho$ and (kinematic) viscosity $\nu$ is electrically conducting and incompressible. Convection is driven by a fixed superadiabatic temperature difference $\Delta T$ between the inner and the outer boundary. Gravity $\mathbf{g}$ increases linearly with radius.

We solve the geodynamo equations in non-dimensional form using the Boussinesq approximation to obtain the velocity field $\mathbf{u}$, magnetic induction $\mathbf{B}$ and temperature perturbation $T$. We adopt $D$ as reference length scale, the viscous diffusion time $D^{2}/\nu$ serves as time unit, temperature is scaled by $\Delta T$ and magnetic induction by $\sqrt{\rho \mu \eta \Omega}$, where $\mu$ is the magnetic permeability and $\eta$ the magnetic diffusivity of the fluid. This results in the following set of equations:
\begin{align}
\begin{split}
   \frac{\partial \mathbf{u}}{\partial t}+\mathbf{u}\cdot \nabla \mathbf{u} + \frac{2}{E} \mathbf{e}_{z}\times \mathbf{u} = &- \nabla P + \frac{Ra}{Pr}\frac{\mathbf{r}}{r_{o}}T + \nabla^{2}\mathbf{u} \ + \\ & + \frac{1}{E Pm}\left(\nabla \times \mathbf{B} \right)\times \mathbf{B}, 
\end{split}\\[5pt]
   \frac{\partial T}{\partial t} + \mathbf{u} \cdot \nabla T =& \ \frac{1}{Pr}\nabla^{2}T,\\[4pt]
   \frac{\partial \mathbf{B}}{\partial t} =& \ \nabla \times \left( \mathbf{u}\times \mathbf{B} \right) + \frac{1}{Pm} \nabla^{2}\mathbf{B} ,\\[4pt]
   \nabla \cdot \mathbf{u}  =& \ 0,\\[5pt]
   \nabla \cdot \mathbf{B} =& \ 0,
 \end{align}
where $P$ is the pressure. The non-dimensional control parameters of the system are the Ekman number
\begin{align}
    E = \frac{\nu}{\Omega D^{2}},
\end{align}
the hydrodynamic Prandtl number
\begin{align}
    Pr = \frac{\nu}{\kappa},
\end{align}
the magnetic Prandtl number
\begin{align}
    Pm = \frac{\nu}{\eta},
\end{align}
and the Rayleigh number
\begin{align}
    Ra = \frac{\alpha g_{o} D^{3} \Delta T}{\nu \kappa},
\end{align}
where $\kappa$ is the thermal diffusivity, $\alpha$ the thermal expansion coefficient and $g_o$ the gravity at the outer boundary.

Both boundaries are assumed to be electrically insulating with vanishing velocity field (no-slip) and fixed temperature. All models are simulated with the open-source numerical code MagIC \citep[][ \url{https://github.com/magic-sph/magic}]{wicht_2002, gastine_etal_2016}, which uses Chebychev polynomials in the radial direction and spherical harmonic decomposition in the angular directions. MagIC relies on the library SHTns \citep[][ \url{https://bitbucket.org/nschaeff/shtns}]{schaeffer_2013} for efficient calculation of the spherical harmonic transforms. Diffusion terms are integrated implicitly in time using a Crank-Nicolson scheme, while a second-order Adams-Bashforth scheme is employed for the explicit treatment of the remaining terms. For the explicit time stepping, numerical stability requires the maximum allowable time step to satisfy a Courant criterion, which is constrained by the spacing of the radial grid points \citep{christensen_etal_1999}. To alleviate the time step restrictions due to the Alfv\'en waves propagating close to the boundaries, we adopt a mapping of the Gauss-Lobatto collocation grid points \citep{kosloff_and_talezer_1993}. This leads to an increase of the maximum allowable time step by up to a factor two and therefore results in a significant reduction of the computational costs \citep[e.g.,][Section~16.9]{boyd_2001}.

\begin{figure*}
\centering
\includegraphics[width=\textwidth]{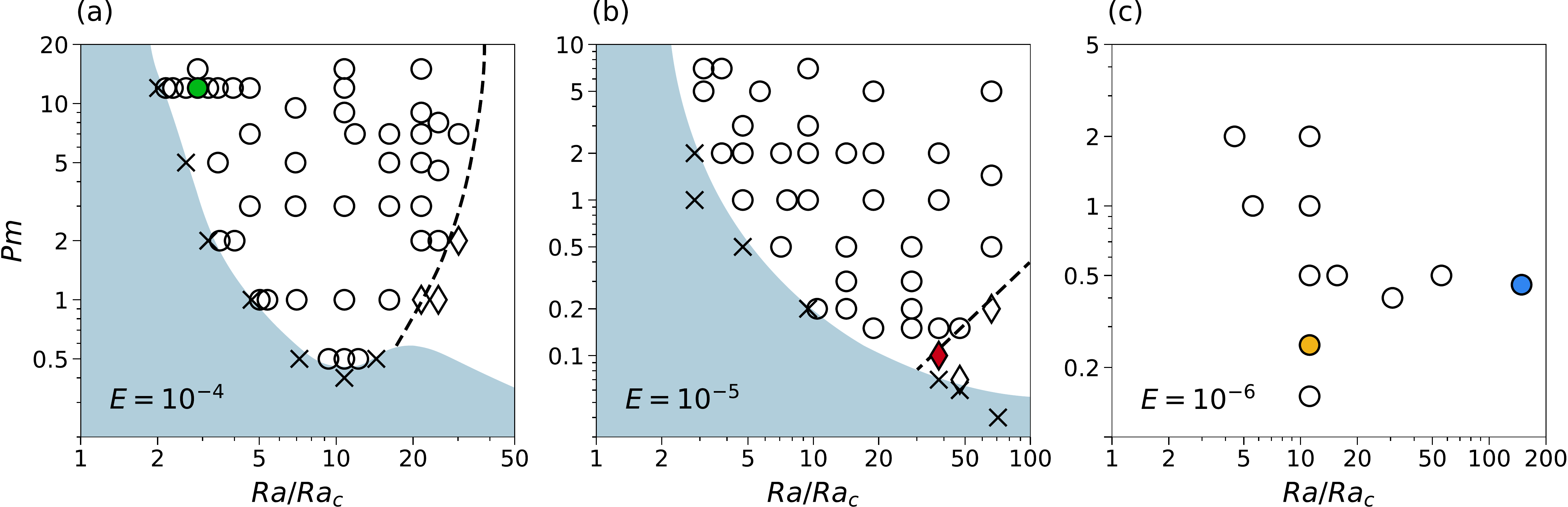}
\caption[]{Regime diagrams for $E = 10^{-4}$, $E = 10^{-5}$ and $E = 10^{-6}$. Shaded regions represent areas of the parameter space where no self-sustained dynamos exist (computationally too expensive to determine for $E = 10^{-6}$). All dynamo models have been computed with $Pr=1$. Circles represent dipolar, diamonds multipolar and crosses failed dynamos. The dashed lines tentatively delineate the transition between dipolar and multipolar dynamos. The control parameters of the dynamo models highlighted by coloured symbols are given in Table \ref{table_parameters}.}
\label{regime_diagrams}
\end{figure*}

To systematically study the force balance that drives the convection in geodynamo models, we perform a series of 95 numerical simulations spanning the parameter range $10^{-6} \leq E \leq 10^{-4}$, $0.07 \leq Pm \leq 15$ and $1.5\times10^{6} \leq Ra \leq 2.66\times10^{10}$ with $Pr = 1$. Fig. \ref{regime_diagrams} shows the regime diagrams for the three different Ekman numbers considered here following \citet{christensen_and_aubert_2006}. The shaded regions represent areas of the parameter space where no self-sustained dynamos could be found. Depending on the geometry of the generated magnetic field we distinguish between dipolar and multipolar dynamos. The transition between these two regimes appears to shift towards higher supercriticalities ($Ra/Ra_{c}$) as the Ekman number decreases, while the onset of dynamo action seems to remain approximately at constant $Ra/Ra_{c}$. Decreasing the Ekman number also extends the region of dynamo action (region of self-sustained dynamos) towards lower $Pm$ \citep{christensen_etal_1999, christensen_and_aubert_2006}. None of the investigated dynamo models in the dipolar regime exhibited reversals of the  the magnetic field polarity. However, for our models with strong magnetic turbulence we expect that reversals could occur, provided the simulations would cover long enough timescales \citep[e.g.,][]{heimpel_and_evans_2013}. For the explored control parameters, verifying this would be extremely demanding in terms of computational resources. Therefore, this is currently not feasible within the scope of a systematic parameter space survey.
To reduce the duration of transients after the start of the simulations, the dynamo models were initiated with an equilibrated solution with similar input parameters whenever possible.
Note that close to the transition of the dipolar to the multipolar regime, bistable dynamos can be found \citep{petitdemange_2018}. This can be attributed to the strength of the seed magnetic field. Similarly, some models close to the onset of dynamo action do require a strong magnetic field at the outset for convection to be able to sustain it. Such bistabilities that depend on the initial conditions have been studied in detail by \citet{petitdemange_2018} for the same physical setup.
In our study, however, we only consider the dipole-dominated dynamos in all of these cases. To obtain an extensive picture of the evolution of the force balance when changing the control parameters, we reproduced several recently published dynamo models \citep{yadav_etal_2016, aubert_etal_2017, dormy_etal_2018}, as well as models covering the parameter space that has been classically explored \citep[e.g.,][]{kutzner_and_christensen_2002, christensen_and_aubert_2006}.

For our following analysis we will mainly focus on four cases which we consider to be representative for the investigated parameter space. Table \ref{table_parameters} summarises the control parameters of these models, along with the coloured symbols used to locate them in the regime diagrams presented in Fig. \ref{regime_diagrams}.

\begin{table}
\caption{Control parameters of four representative dynamo models with coloured symbols to locate them in the regime diagrams (Fig. \ref{regime_diagrams}).}
\centering
\begin{tabular}{cccccc}
\hline
\multicolumn{2}{c}{Model} & $E$ & $Ra$ & $Ra/Ra_{c}$ & $Pm$ \\ \hline
\hspace{1.1mm} \includegraphics[width=2.1mm, height=2.1mm]{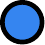} & \textbf{A} \hspace{1.1mm} & $10^{-6}$ & $2.66 \times 10^{10}$ & $148.5$ & $0.456$   \\[0.5mm]
\hspace{1.1mm} \includegraphics[width=2.1mm, height=2.1mm]{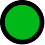} & \textbf{B} \hspace{1.1mm} & $10^{-4}$ & $2.2 \times 10^{6}$ & $3.2$ & $12$  \\[0.5mm]
\hspace{1.1mm} \includegraphics[width=2.1mm, height=2.1mm]{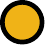} & \textbf{C} \hspace{1.1mm} & $10^{-6}$ & $2 \times 10^{9}$ & $11.2$ & $0.25$ \\
\hspace{1.1mm} \includegraphics[width=1.8mm, height=3.0mm]{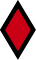} & \textbf{D} \hspace{1.1mm} & $10^{-5}$ & $4 \times 10^{8}$ & $37.8$ & $0.1$ \\ \hline
\end{tabular}
\label{table_parameters}
\end{table}

\section{Results}
\subsection{Force balance spectra}
\begin{figure*}
\centering
\includegraphics[width=0.9\textwidth]{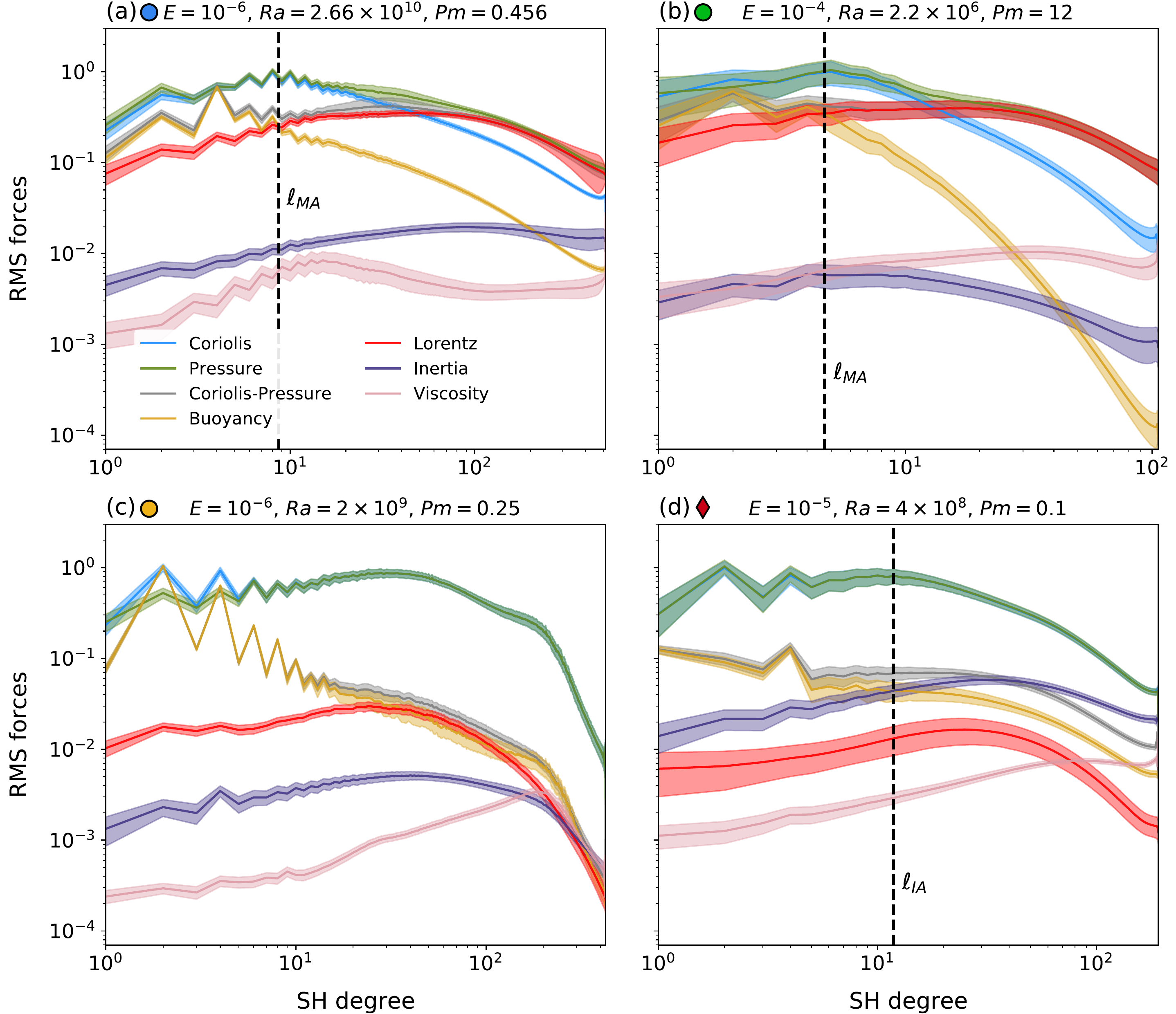}
\caption[]{Force balance spectra for examples of different types of force balances excluding the viscous boundary layers. The spherical harmonic contributions of the r.m.s. forces are normalised with respect to the peak of the Coriolis force. The solid lines represent the time averages of the forces. The corresponding shaded regions represent one standard deviation. a-b: Examples of QG-MAC balances of dipole-dominated dynamos at $E=10^{-6}$ and $E=10^{-4}$. c: Example of a special case with control parameters close to the onset of dynamo action at $E=10^{-6}$. d: Example of QG-CIA balance of a multipolar dynamo at $E=10^{-5}$. The four coloured symbols refer to the location of the dynamos in the parameter space (see Fig. \ref{regime_diagrams}). The vertical dashed lines correspond to the cross-over length scales defined in section \ref{section_crossover}.}
\label{force_balances}
\end{figure*}

For our systematic study of the force balance, we follow the method introduced by \citet{aubert_etal_2017} and decompose each force into spherical harmonic contributions:
\begin{align}
F^{2}_{rms} = \frac{1}{V} \int_{r_{i}+b}^{r_{o}-b} \sum_{\ell=0}^{\ell_{\mathrm{max}}} \sum_{m=0}^{\ell} F_{\ell m}^{2} r^{2} \mathrm{d}r = \sum_{\ell=0}^{\ell_{\mathrm{max}}} F_{\ell}^{2},
\label{force_integration}
\end{align}
where $b$ represents the thickness of the viscous boundary layers. Viscous boundary layers are excluded from the calculations since we are primarily interested in the force balance in the bulk of the fluid. The resulting force balance spectra of the four cases given in Table \ref{table_parameters} are illustrated in Fig. \ref{force_balances}. Most force balance spectra of the dynamo models in the dipolar regime are structurally very similar.
The model \textbf{A} (see Fig. \ref{force_balances}a), which is among the ``path''-dynamos analysed by \citet{aubert_etal_2017}, can be considered as a typical example. At zeroth order, it is characterised by a quasi-geostrophic balance between pressure and the Coriolis force. The ageostrophic part of the Coriolis force is then balanced by buoyancy at small spherical harmonic degrees and by the Lorentz force at large $\ell$. This QG-MAC balance has been identified in several recent studies \citep[e.g.,][]{yadav_etal_2016, schaeffer_etal_2017, aubert_etal_2017}. This balance is, however, quite significantly disturbed by inertia and viscous forces since they are only one to two orders of magnitude smaller than the leading-order forces. 

\citet{dormy_2016} recently suggested that dynamos governed by a magnetostrophic balance at zeroth order can be attained even in a computationally moderate regime by adopting a setup close to the onset of convection to minimise inertial effects and with large $Pm$ to maintain a strong influence of the Lorentz force. Our study confirms that in these dynamos the Lorentz force is of approximately the same magnitude as the total Coriolis force. Therefore, these models do indeed approach magnetostrophy when considering volume-integrated forces. However, the length scale dependent analysis of model \textbf{B} \citep[which corresponds to one of the configurations considered by][]{dormy_etal_2018} using the force balance spectra (see Fig. \ref{force_balances}b) reveals the same basic structure as for the QG-MAC cases, i.e. like for model \textbf{A} we observe a geostrophic balance at zeroth order, followed by a first-order balance between the ageostrophic Coriolis, buoyancy and Lorentz forces. This indicates that these models do not represent a force balance regime that is different from the one of most dipole-dominated dynamos.
Yet, due to the role of inertia getting minimised, the separation between the Lorentz force and second-order forces increases compared to QG-MAC cases at larger supercriticalities. Therefore, one may refer to such dynamo models as strong-field cases. 

The only occurrences of dipole-dominated dynamos that cannot be attributed to the QG-MAC regime can be found in regions of the parameter space close to the onset of dynamo action. Model \textbf{C} (see Fig. \ref{force_balances}c) can be considered as an example for such dynamos at low $Pm$, which are characterised by a significantly weaker Lorentz force compared to typical QG-MAC cases. While the Lorentz force is still larger than inertia and viscous forces at large scales, it becomes very weak towards smaller scales and as a result does not balance the ageostrophic Coriolis force at any point. 

By increasing the vigour of convection one eventually reaches the transition from the dipolar to the multipolar regime \citep[e.g.,][]{kutzner_and_christensen_2002, christensen_and_aubert_2006}. Model \textbf{D} (see Fig. \ref{force_balances}d) is an example of a multipolar dynamo close to this transition. Its force balance spectrum features a significantly weaker Lorentz force than the QG-MAC cases. This decrease of the Lorentz force might be related to the increasing role of inertia \citep{christensen_and_aubert_2006} or the breaking of the equatorial symmetry of the flow \citep{garcia_etal_2017}. For multipolar cases inertia becomes a first-order contribution to the force balance, such that they are controlled by a first-order CIA (Coriolis-Inertia-Archimedean) balance \citep{gillet_and_jones_2006}. At zeroth order these dynamos exhibit a quasi-geostrophic balance again. Therefore, analogously to QG-MAC, we refer to this type of force balance as QG-CIA balance.

\subsubsection{``Strong-fieldness"}
\begin{figure*}
\centering
\includegraphics[width=\textwidth]{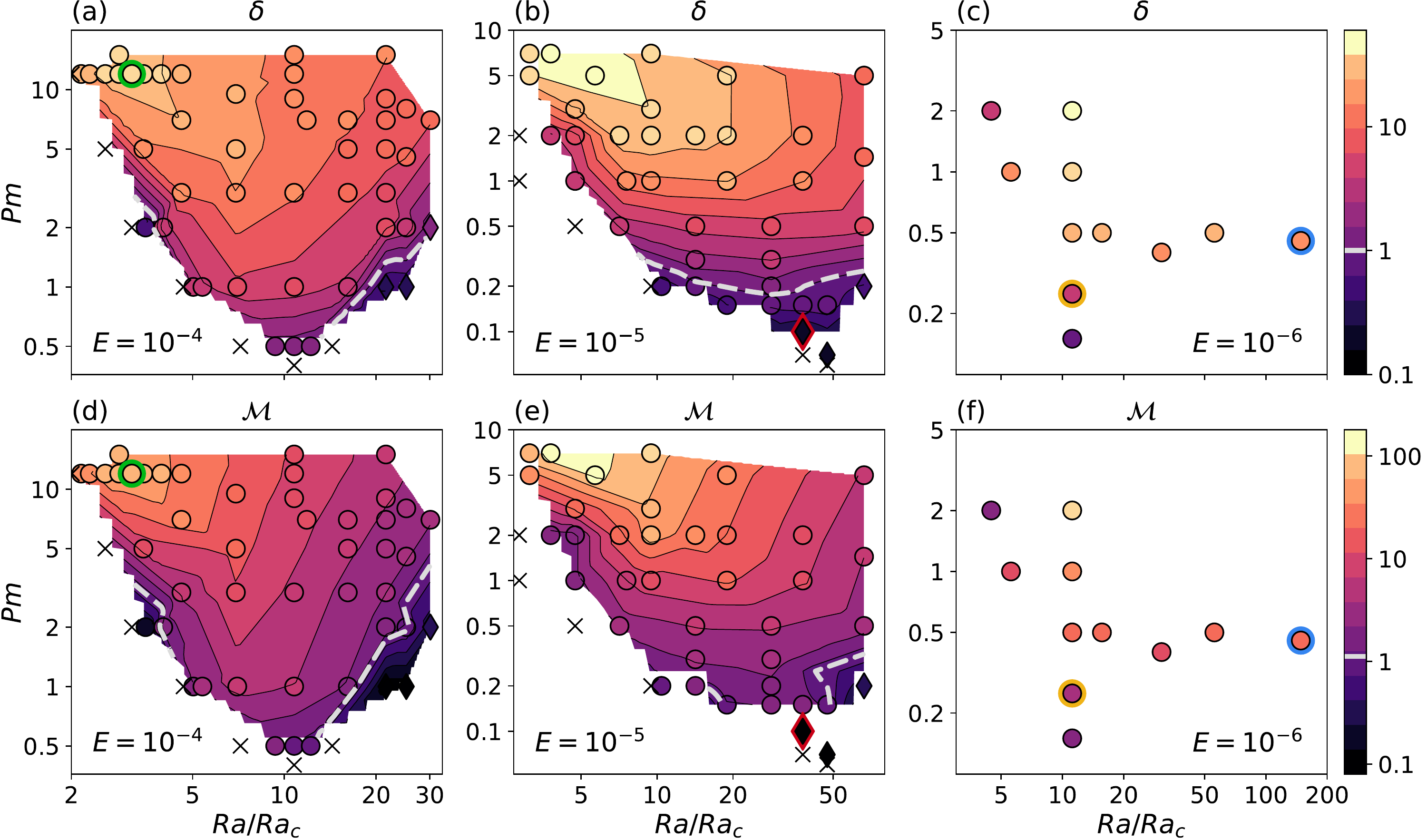}
\caption[]{Regime diagrams for Ekman numbers $E=10^{-4}$, $E=10^{-5}$ and $E=10^{-6}$ with linear interpolations of the ``strong-fieldness" $\delta$ (Eq. \ref{delta_eqn}, panels a-c) and the ratio of the magnetic and kinetic energies $\mathcal{M}$ (Eq. \ref{M_eqn}, panels e-f). 
The light grey dashed lines correspond to a value of 1. The symbols are filled with the values computed from the simulation output. The meaning of the symbols is the same as in Fig. \ref{regime_diagrams}. The symbols of the models given in Table \ref{table_parameters} are highlighted by coloured edges.}
\label{strongfieldness_energyratio}
\end{figure*}
To evaluate, based on the force balance spectra, which of the dynamos can be attributed to the strong-field regime, we introduce the ``strong-fieldness" $\delta$. This parameter quantifies the separation between the Lorentz force and the second-order forces (inertia and viscous forces) and is defined as:
\begin{align}
    \delta = \sqrt{\frac{\sum_{\ell=1}^{\ell_{\mathrm{max}}} F_{\mathrm{Lorentz,}\ell}^{2}}{\sum_{\ell=1}^{\ell_{\mathrm{max}}} \max \left(F_{\mathrm{inertia,}\ell}^{2}, F_{\mathrm{viscous,}\ell}^{2} \right)}}.
\label{delta_eqn}
\end{align}
Fig. 3a-b show extrapolated contour levels of $\delta$ for the investigated parameter space at $E = 10^{-4}$ and $E = 10^{-5}$. For $E=10^{-6}$, the limited number of simulations does not allow a meaningful linear interpolation required to draw the contour lines. As a consequence, only the data points are displayed as a scatterplot in Fig. \ref{strongfieldness_energyratio}c. We observe that $\delta$ reaches its maximum for dynamos at low $Ra/Ra_{c}$ and high $Pm$ for all three Ekman numbers. This confirms the results by \citet{dormy_2016} and \citet{dormy_etal_2018} who suggested that strong-field dynamos can be attained for this parameter range. The smallest values of $\delta$ are found close to the onset of dynamo action and in the multipolar regime where the Lorentz force falls below the level of inertia.
Decreasing the Ekman number from $E = 10^{-4}$ to $E = 10^{-6}$ leads to an overall increase of $\delta$. In parallel, the parameter region of dynamos with $\delta \gg 1$, which corresponds to QG-MAC dynamos, gradually extends towards lower values of $Pm$.

The influence of viscous forces on $\delta$ decreases strongly with decreasing Ekman number. As a consequence, $\delta$ can be approximated by the ratio between the magnetic and kinetic energies $\mathcal{M}$:
\begin{align}
\mathcal{M} = \frac{E_{\mathrm{mag}}}{E_{\mathrm{kin}}} = \frac{B^{2}}{\rho \mu U^{2}},
\label{M_eqn}
\end{align}
which represents a proxy for the relative magnitudes of the Lorentz force and inertia. The linear interpolations of $\mathcal{M}$ for the explored parameter space are shown in Fig. \ref{strongfieldness_energyratio}d-f. Comparison of the integral diagnostic $\mathcal{M}$ to $\delta$ shows a broad agreement. The discrepancy in the amplitude between the two parameters can be explained by the independence of $\mathcal{M}$ on length scales, while they are inherently included in the definiton of $\delta$ due to the explicit calculation of the forces. Additionally, viscosity still represents a sizeable contribution at large Ekman numbers.

\subsubsection{Cross-over length scale}
\label{section_crossover}
Following \citet{aubert_etal_2017} we use the spectral representations of the forces to introduce the cross-over length scales $d_{\perp}$. These are defined as the length scales where two forces are of equal amplitude. This implies that these forces are in balance under the constraint of some remainder of the Coriolis force. Hence, this scale corresponds in fact to a three-terms balance. To obtain the cross-over length scale which corresponds to the first-order force balance, we therefore determine the spherical harmonic degree, $\ell_{\mathrm{MA}}$, where buoyancy and the Lorentz force are of equal magnitude  in the case of QG-MAC dynamos. Analogously, we also identify the cross-overs between buoyancy and inertia, $\ell_{\mathrm{IA}}$, and buoyancy and viscous forces, $\ell_{\mathrm{VA}}$, for QG-CIA and QG-VAC (Quasi-Geostrophic Viscous-Archimedean-Coriolis) dynamos, respectively. Hence, we determine the three following spherical harmonic degrees:
\begin{align}
\ell_{\mathrm{MA}} &= \min_{\ell} \left( \vert F_{\mathrm{Lorentz,}\ell} - F_{\mathrm{buoyancy,}\ell} \vert \right), \\
\ell_{\mathrm{IA}} &= \min_{\ell} \left( \vert F_{\mathrm{inertia,}\ell} - F_{\mathrm{buoyancy,}\ell} \vert \right), \\
\ell_{\mathrm{VA}} &= \min_{\ell} \left( \vert F_{\mathrm{viscous,}\ell} - F_{\mathrm{buoyancy,}\ell} \vert \right).
\end{align}
For our example cases (see Table \ref{table_parameters}), the cross-overs are highlighted by vertical dashed lines in Fig. \ref{force_balances}. The associated length scales are defined as $d_{\perp}=\pi/\ell_{\perp}$. Note that none of the investigated models exhibits a first-order QG-VAC balance since we exclude viscous boundary layers from the integration of the forces and the bulk viscosity is too small to enter the first-order force balance. Several dynamos close to the onset of dynamo action feature force balances where the Lorentz force, inertia and buoyancy are of the same order of magnitude. Due to the lack of separation between the three forces, it does not make sense to define a cross-over length scale based on one individual force in such cases. Dynamos featuring a force balance with ill-posed cross-overs, e.g. model \textbf{C} (see Fig. \ref{force_balances}c), also do not allow the determination of a unique relevant cross-over length scale.

\subsection{Convective pattern}
\begin{figure*}
\centering
\includegraphics[width=0.75\textwidth]{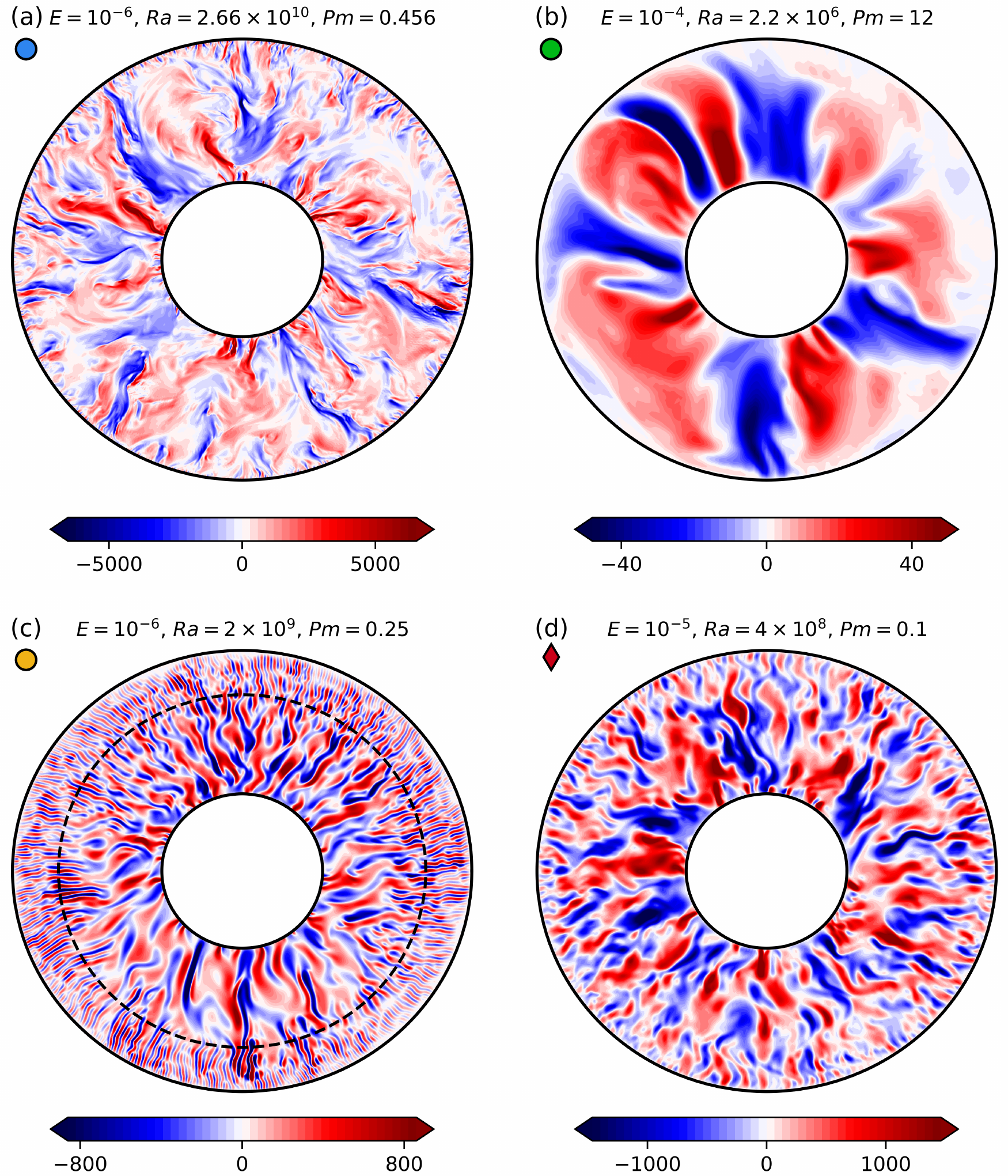}
\caption[]{Radial velocity in the equatorial plane for the four simulations highlighted in Fig. \ref{regime_diagrams}. The abrupt change of the convective length scale towards the outer boundary in the equatorial plane of model \textbf{C} (panel c) is highlighted by a dashed circle at radius $r \sim 0.8 \ r_{o}$.}
\label{planforms}
\end{figure*}

To qualitatively analyse the effect of the governing force balance on the dominant length scale of the convective flow, we turn to the equatorial planes of the radial velocity of the dynamos, which are presented in Fig. \ref{planforms}.

In the equatorial planes of both QG-MAC dynamos (models \textbf{A} and \textbf{B}) one can observe elongated structures of the scale of the system size despite the large difference in the input parameters (see Fig. \ref{planforms}a-b). Since model \textbf{A} is far more supercrititcal than model \textbf{B}, it also shows a greater range of length scales with small-scale features developing close to the outer boundary \citep[e.g.,][]{sakuraba_and_roberts_2009}. The equatorial plane of model \textbf{C} (see Fig. \ref{planforms}c) features a rather abrupt change in the size of the convective cells as it transitions from large elongated structures in the interior of the shell to very small scales towards the outer boundary. \citet{yadav_etal_2016} also observed such layers of small-scale convection in their simulations at $E = 10^{-6}$, which they attributed to a weak Lorentz force in these regions. The equatorial plane of the multipolar dynamo (model \textbf{D}, see Fig. \ref{planforms}d) shows mostly small convective scales, which resemble that of non-magnetic convection due to inertia being much stronger compared to the Lorentz force (see Fig. \ref{force_balances}d) \citep[e.g.,][]{gillet_and_jones_2006, gastine_etal_2016}. A gradual decrease of the convective scale with radius can also be observed in the equatorial plane of the multipolar dynamo.

The dominant length scales of these models can be characterised by the peaks of the poloidal kinetic energy spectra:
\begin{equation}
\ell_{\mathrm{pol}} = \max_{\ell} \left( E_{\mathrm{pol}} \right).
\end{equation}
Comparison of the cross-over length scales to the observed number of up- and downwellings, as well as to $\ell_{\mathrm{pol}}$, shows a satisfactory agreement for the QG-MAC cases (models \textbf{A} and \textbf{B}). This becomes, however, more challenging for models \textbf{C} and \textbf{D} due to the overall smaller convective cells and the additional strong radial dependence of the length scales. This suggests that analysing the radial dependence of the force balances will help to fully understand how the cross-over length scales relate to the observed convective scales.

\subsection{2D force spectra}
\begin{figure*}
\centering
\includegraphics[width=\textwidth]{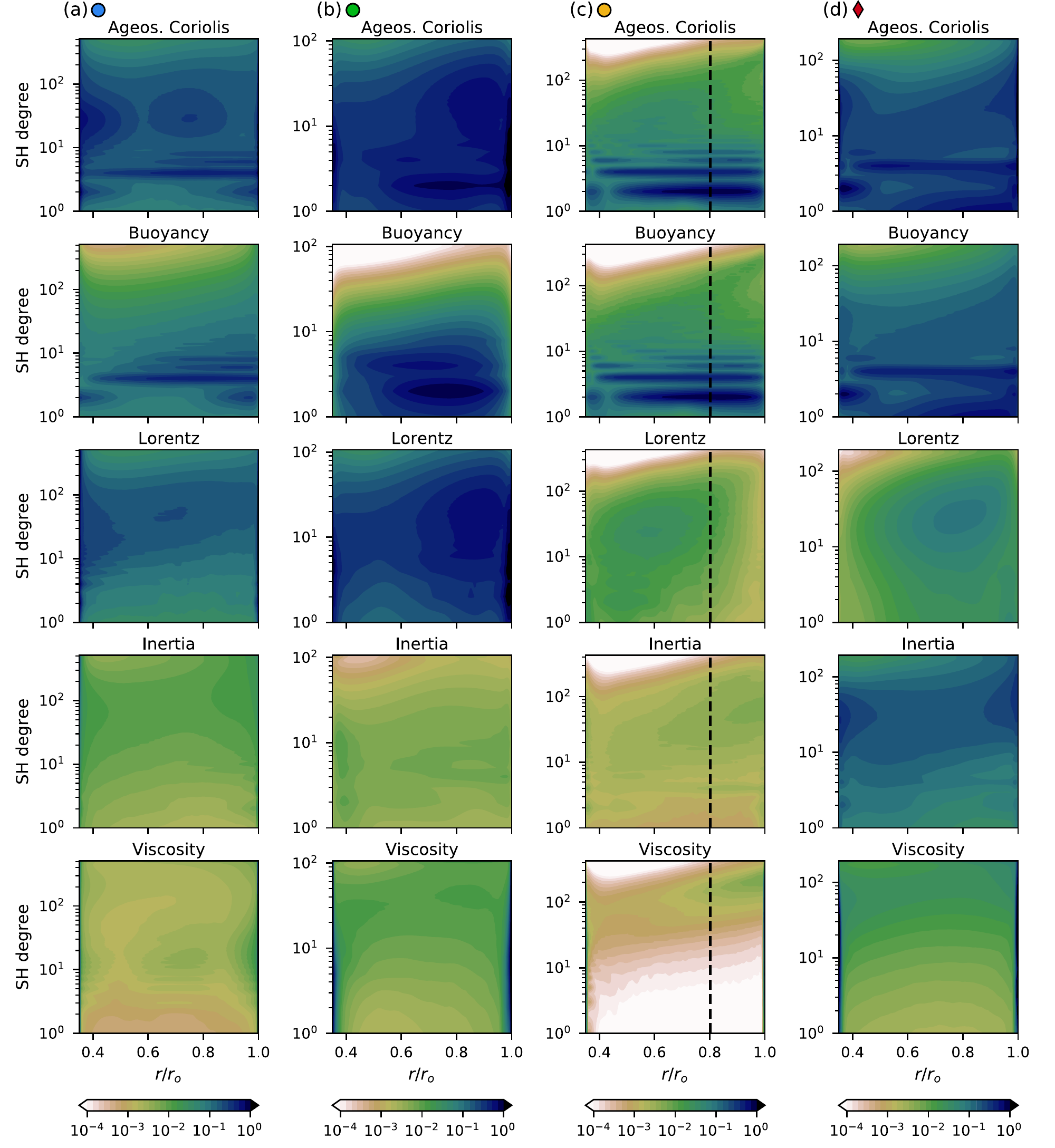}
\caption[]{2D force balance spectra of the dynamo models highlighted in Fig. \ref{regime_diagrams}. The geostrophic balance formed by pressure and Coriolis force is omitted since it is on a global scale present in all of the investigated dynamo models. The 2D force spectra of each model are normalised by the maximum of the forces excluding the viscous boundary layers. The vertical dashed lines at $r \sim 0.8 \ r_{o} $ in panel (c) correspond to the change of the convective length scale that can be observed in the equatorial plane of model \textbf{C} (see Fig. \ref{planforms}c).}
\label{l_vs_r_forces}
\end{figure*}
Changes of the convective length scale in the radial direction, like in the equatorial plane of case \textbf{C} (see Fig. \ref{planforms}c), suggest an underlying change in the dominant force balance with radius. To quantify this effect, we introduce a measure of the local forces defined as:
\begin{align}
F_{\ell}^{2}\left( r \right) = \sum_{m=0}^{\ell} F_{\ell m}^{2}.
\end{align} 
The resulting 2D force spectra of the dynamo models discussed above are illustrated in Fig. \ref{l_vs_r_forces}. We exclude pressure and Coriolis forces from this representation since all investigated models are, on global scales, quasi-geostrophic at zeroth order, and restrict our focus on the contributions of first- and second-order forces.

The 2D spectra of the QG-MAC dynamos (models \textbf{A} and \textbf{B}, see Fig. \ref{l_vs_r_forces}a-b) show a balance between the ageostrophic Coriolis force and buoyancy on large scales (small spherical harmonic degrees) and by the Lorentz force on small scales (large $\ell$) for all radii (excluding viscous boundary layers). In model \textbf{A} the Lorentz force is very strong throughout the entire volume. Yet, one can observe a maximum close to the inner boundary from which it tends to decrease with increasing radius. 
Model \textbf{B} also displays a strong Lorentz force throughout the entire shell, however, with a localised maximum towards the outer boundary at intermediate length scales. Note that buoyancy only slightly depends on radius in model \textbf{A}, while in model \textbf{B} it is significantly larger in the inner part of the volume compared to the boundaries. This is expected as the lower vigour of convection in models with Rayleigh numbers close to the onset of convection leads to a overall less efficient heat transport, and therefore to the formation of thick thermal boundary layers.

\begin{figure*}
\centering
\includegraphics[width=\textwidth]{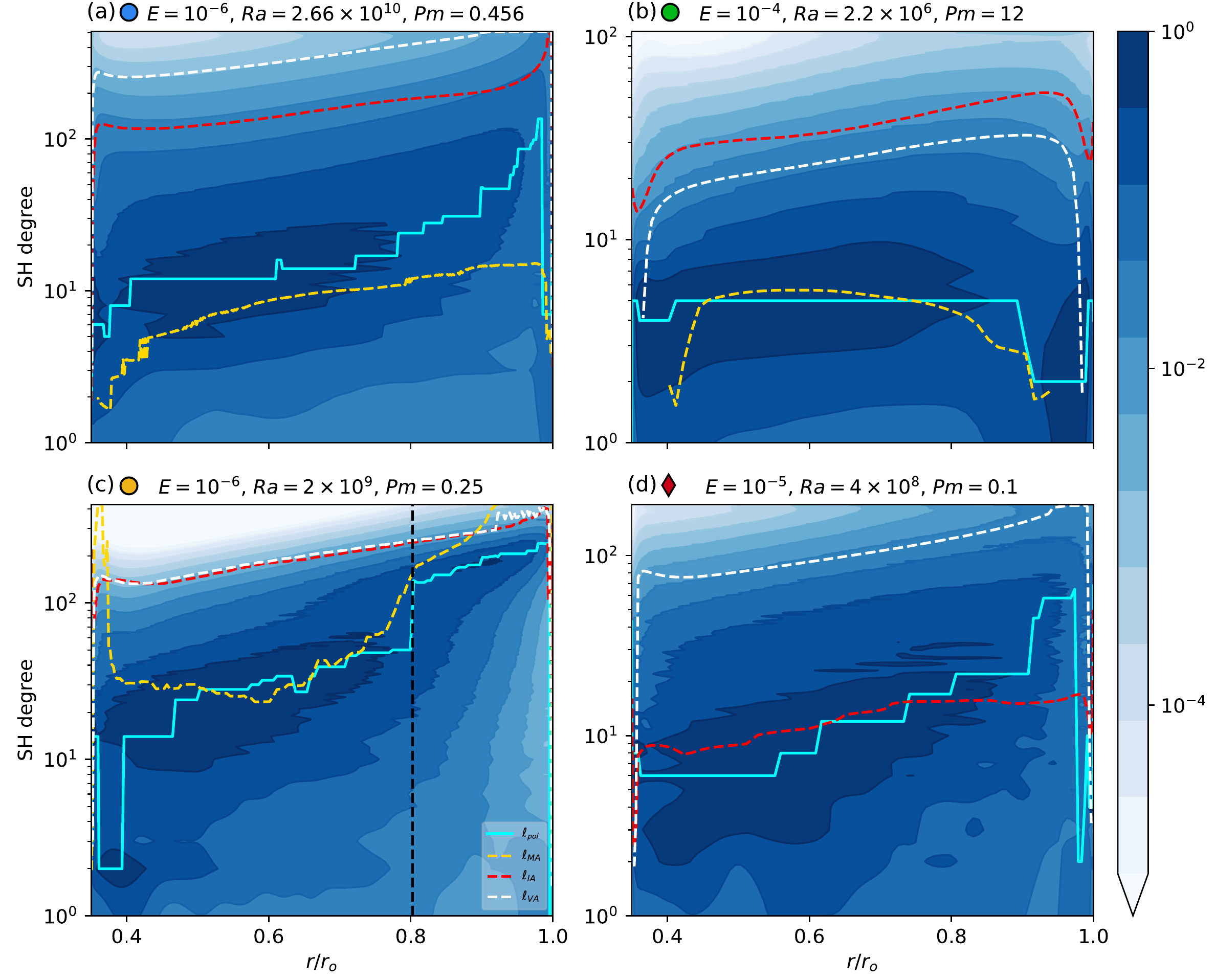}
\caption[]{2D-spectra of the poloidal kinetic energy of the dynamo models highlighted in Fig. \ref{regime_diagrams}. Each 2D-spectrum is normalised by its maximum value. The solid blue line connects the peaks of the poloidal kinetic energy spectra ($\ell_{\mathrm{pol}}$) for each each radial level. The dashed lines are the harmonic degrees of the crossings in the corresponding force balance spectra; yellow: $\ell_{\text{MA}}$, scale at which buoyancy and Lorentz force are equal; red: $\ell_{\text{IA}}$, scale at which buoyancy and inertia are equal; grey: $\ell_{\text{VA}}$, scale at which buoyancy and viscous forces are equal. The vertical dashed line at $r \sim 0.8 \ r_{o} $ in panel (c) corresponds to the change of the convective length scale that can be observed in the equatorial plane of model \textbf{C} (see Fig. \ref{planforms}c).}
\label{ekin_pol_lengthscales}
\end{figure*}

In the 2D force spectra of model \textbf{C} (see Fig. \ref{l_vs_r_forces}c), the ageostrophic Coriolis force is balanced by buoyancy and Lorentz force for radii $r \lesssim 0.8 \ r_{o} $, therefore exhibiting a first-order QG-MAC balance which explains the elongated flow structures in this portion of the volume. However, the Lorentz force is overall considerably weaker on small length scales compared to most QG-MAC dynamos. For larger radii, we observe a significant decrease of the Lorentz force. Additionally, inertia and viscosity increase towards the outer boundary. As a consequence, the ageostrophic Coriolis force is in the outer part of the shell almost entirely balanced by buoyancy and to a smaller extent by inertia and viscous forces. Since inertia and viscous forces are slightly larger than the Lorentz force in this region, the force balance of the dynamo is close to a QG-CIA/QG-VAC regime which appears to be the reason for the layer of small-scale convection that is visible in Fig. \ref{planforms}c. This change in the governing force balance depending on the radius also explains why the cross-over length scale in the fully integrated force balance spectrum is ill-posed for model \textbf{C} (see Fig. \ref{force_balances}c).

In the multipolar dynamo (model \textbf{D}, see Fig. \ref{l_vs_r_forces}d), the ageostrophic Coriolis force is largely balanced by buoyancy and inertia at all radii. The Lorentz force remains smaller than inertia throughout the entire shell, and therefore does never contribute to the first-order force balance. As a consequence, the dynamo is controlled by a QG-CIA balance throughout the entire shell.

The 2D force spectra of all four models show that the viscosity is predominantly confined to the inner and outer boundaries, which justifies our decision to exclude the viscous boundary layers when calculating the integrated forces (see Eq. \ref{force_integration}). While at $E=10^{-4}$ the viscous boundary layers still extend into the shell, they become very thin at lower Ekman numbers like $E=10^{-6}$. At $E = 10^{-4}$, both the bulk viscosity and inertia are only about one order of magnitude smaller than the leading order forces. Hence, they may still significantly influence the overall force balance. Decreasing the Ekman number to $E=10^{-6}$ leads to the bulk viscosity and inertia being two to three orders of magnitude lower than the first-order forces.

\begin{figure*}
\centering
\includegraphics[width=\textwidth]{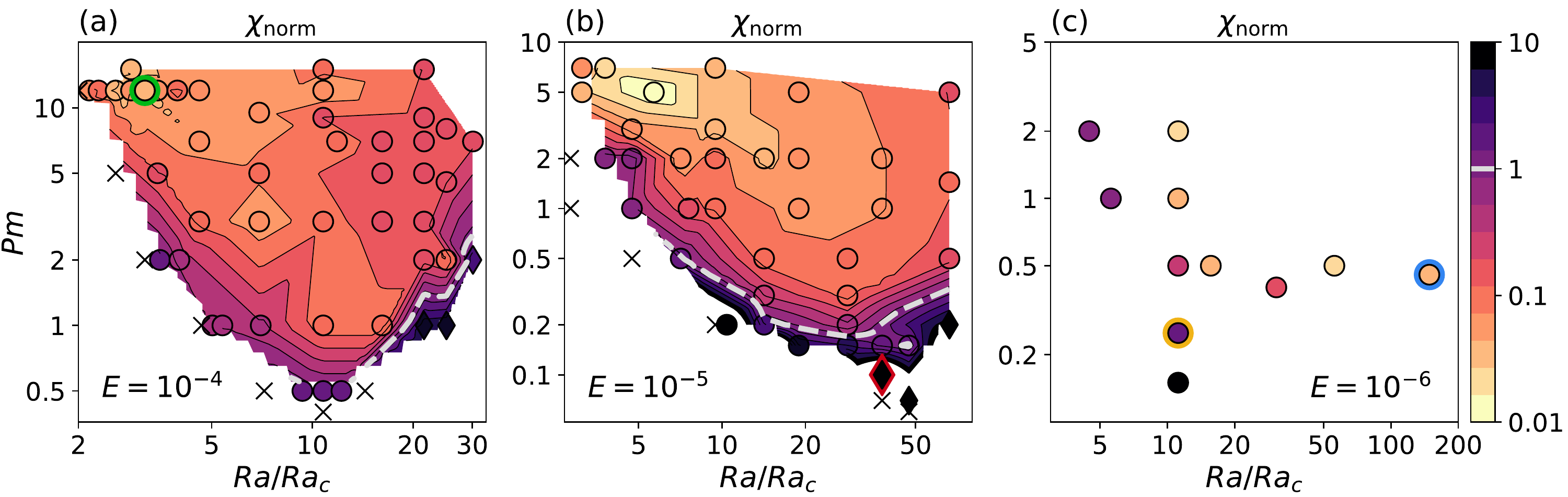}
\caption[]{Regime diagrams for Ekman numbers $E=10^{-4}$, $E = 10^{-5}$ and $E = 10^{-6}$ with linear interpolations of $\chi_{\mathrm{norm}}$ (Eq. \ref{chinorm}), with small values of $\chi_{\mathrm{norm}}$ corresponding to QG-MAC dynamos and large values to non-QG-MAC dynamos. The symbols are filled with the values computed from the simulation outputs. The meaning of the symbols is the same as in Fig. \ref{regime_diagrams}. The symbols of the models given in Table \ref{table_parameters} are highlighted by coloured edges.}
\label{contour_MAC_norm}
\end{figure*}

\subsubsection{Cross-over length scales of 2D force spectra}
The 2D force spectra further allow us to determine the cross-over length scales at each radial level. To obtain an idea on how well these match the convective scale in the equatorial plane, we compare them to $\ell_{\mathrm{pol}} \left( r \right)$. Fig. \ref{ekin_pol_lengthscales} illustrates the 2D spectra of the poloidal kinetic energy for the four example cases, along with the cross-over length scales linked to the different types of force balances for each radial level. While it is generally possible to determine the scales at which the buoyancy-inertia and buoyancy-viscous pairs are of equal magnitude for nearly all radii, the crossing between buoyancy and Lorentz force becomes ill-posed in the thermal boundary layers, where buoyancy is weak. This is especially the case for the strong-field dynamos that are run at low supercritical Rayleigh numbers and are therefore only weakly driven, which leads to the formation of thick thermal boundary layers. In several multipolar dynamos and dynamos close to the onset of convection it is also not possible to determine $\ell_{\text{MA}}$ in some parts of the volume, or in a few cases even the entire shell, because of the Lorentz force being too weak relative to buoyancy.

Fig. \ref{ekin_pol_lengthscales}a-b show that for both QG-MAC dynamos (models \textbf{A} and \textbf{B}) $\ell_{\text{MA}}$ is in good agreement with $\ell_{\mathrm{pol}}$ in the bulk of the volume. In addition, $\ell_{\text{IA}}$ and $\ell_{\text{VA}}$ do not match $\ell_{\mathrm{pol}}$ with the exception of the viscous boundary layers, where viscosity and therefore $\ell_{\text{VA}}$ becomes relevant. The separation between the length scales corresponding to the different force balances is larger in model \textbf{A} compared to model \textbf{B} since it operates at a lower Ekman number.

For model \textbf{C} (see Fig. \ref{ekin_pol_lengthscales}c), $\ell_{\text{MA}}$ is close to $\ell_{\mathrm{pol}}$ for most of the interior of the volume. Although close to the inner boundary it is not possible to determine a relevant crossing as the spectral contributions of the Lorentz force and buoyancy overlap. This corresponds, however, only to a small portion of the total volume. For the outer region $\ell_{\text{MA}}$ starts to deviate from $\ell_{\mathrm{pol}}$ at $r \gtrsim 1.3$. Beyond this radius, $\ell_{\text{IA}}$ and $\ell_{\text{VA}}$ corresponding to a CIA and VAC balance, respectively, start to match the peaks of the poloidal energy better.
This confirms what we observe in the 2D force spectra (see Fig. \ref{l_vs_r_forces}c), which revealed a QG-MAC balance in the interior of the shell and a QG-CIA/QG-VAC regime towards the outer boundary. Hence, the agreement of the different cross-over length scales with $\ell_{\mathrm{pol}}$, i.e. the convective length scales, indeed appears to reflect the force balance at the given radius.

For the multipolar dynamo (model \textbf{D}, see Fig. \ref{ekin_pol_lengthscales}d), $\ell_{\text{IA}}$ fits $\ell_{\mathrm{pol}}$ best. This is expected since the first-order force balance corresponds to a CIA balance. Due to the weak Lorentz force it is not possible to determine $\ell_{\text{MA}}$ for this dynamo.

These results suggest that by quantifying the agreement between the peak of the poloidal kinetic energy spectra and the harmonic degrees of the different types of crossings, we can obtain a measure for the type of force balance that controls the dynamo. Therefore, we calculate the volumetric relative misfit  between $\ell_{\mathrm{pol}}$ and the crossings $\ell_{\perp}$ ($\ell_{\text{MA}}$, $\ell_{\text{IA}}$ and $\ell_{\text{VA}}$) using the following formula:
\begin{align}
\chi_{i} = \sqrt{\frac{4 \pi}{V} \int_{r} \left( \frac{\ell_{\mathrm{pol}}\left( r \right)-\ell_{\perp}\left( r \right)}{\ell_{\mathrm{pol}}\left( r \right)} \right)^{2} r^{2} \mathrm{d}r},
\label{chi}
\end{align}
where $i = \left[ \mathrm{MAC}, \mathrm{CIA}, \mathrm{VAC} \right]$. We again exclude viscous boundary layers. Additionally, we restrict the computation of the misfits of all three types of crossings to the part of the volume where $\ell_{\text{MA}}$ is defined.  Since the focus of our study is on finding dynamo models with a force balance relevant to the Earth's core conditions, a simple classification in QG-MAC and non-QG-MAC dynamos is sufficient at first. Therefore, we normalise the misfit that we obtain for the QG-MAC balance by the one for the next best fitting force balance (either QG-CIA or QG-VAC), i.e.
\begin{align}
\chi_{\mathrm{norm}} = \frac{\chi_{\mathrm{MAC}}}{\mathrm{min}\left(\chi_{\mathrm{CIA}},\chi_{\mathrm{VAC}} \right)}.
\label{chinorm}
\end{align}
This allows us to quantify the discrepancy between the misfit of the cross-over length scales corresponding to a QG-MAC balance and the other types of force balances and therefore essentially whether the dynamo is controlled by a QG-MAC balance or not. Linear interpolations of $\chi_{\mathrm{norm}}$ for the investigated parameter space are shown in Fig. \ref{contour_MAC_norm}. Comparison of Fig. \ref{contour_MAC_norm} to the same type of representation for $\delta$ and $\mathcal{M}$ (see Fig. \ref{strongfieldness_energyratio}) shows that in the region of the parameter space at low $Ra/Ra_{c}$  and high $Pm$, where one can find the strong-field dynamos, we also observe the smallest values for $\chi_{\mathrm{norm}}$, i.e. the best agreement between $\ell_{\mathrm{pol}}$ and $\ell_{\text{MA}}$. One notable difference is that $\delta$ and $\mathcal{M}$ transition generally quite smoothly throughout the parameter space from large to small values as indicated by the equidistant spacing of the contour lines. For $\chi_{\mathrm{norm}}$, however, the smaller values found in the region of strong-field dynamos extend to lower values of $Pm$, before decaying rather steeply near the onset of dynamo action and close to the transition from the dipolar to the multipolar regime. Therefore, $\chi_{\mathrm{norm}}$ seems to allow for a better distinction between the different force balance regimes. Decreasing the Ekman number gradually extends the boundaries of the QG-MAC regime towards lower $Pm$ and higher $Ra/Ra_{c}$.

\section{Discussion}
Dynamo action in the Earth's outer core is expected to be controlled by a balance between pressure, Coriolis, buoyancy and Lorentz forces, with marginal contributions from inertia and viscous forces \citep[e.g.,][]{roberts_and_king_2013}. Current numerical simulations of the geodynamo, however, operate at much larger inertia and viscosity because of computational limitations. This has lead to conflicting interpretations of the classical data set, casting some doubt on the physical relevance of these models.

\begin{figure}
\centering
\includegraphics[width=0.44\textwidth]{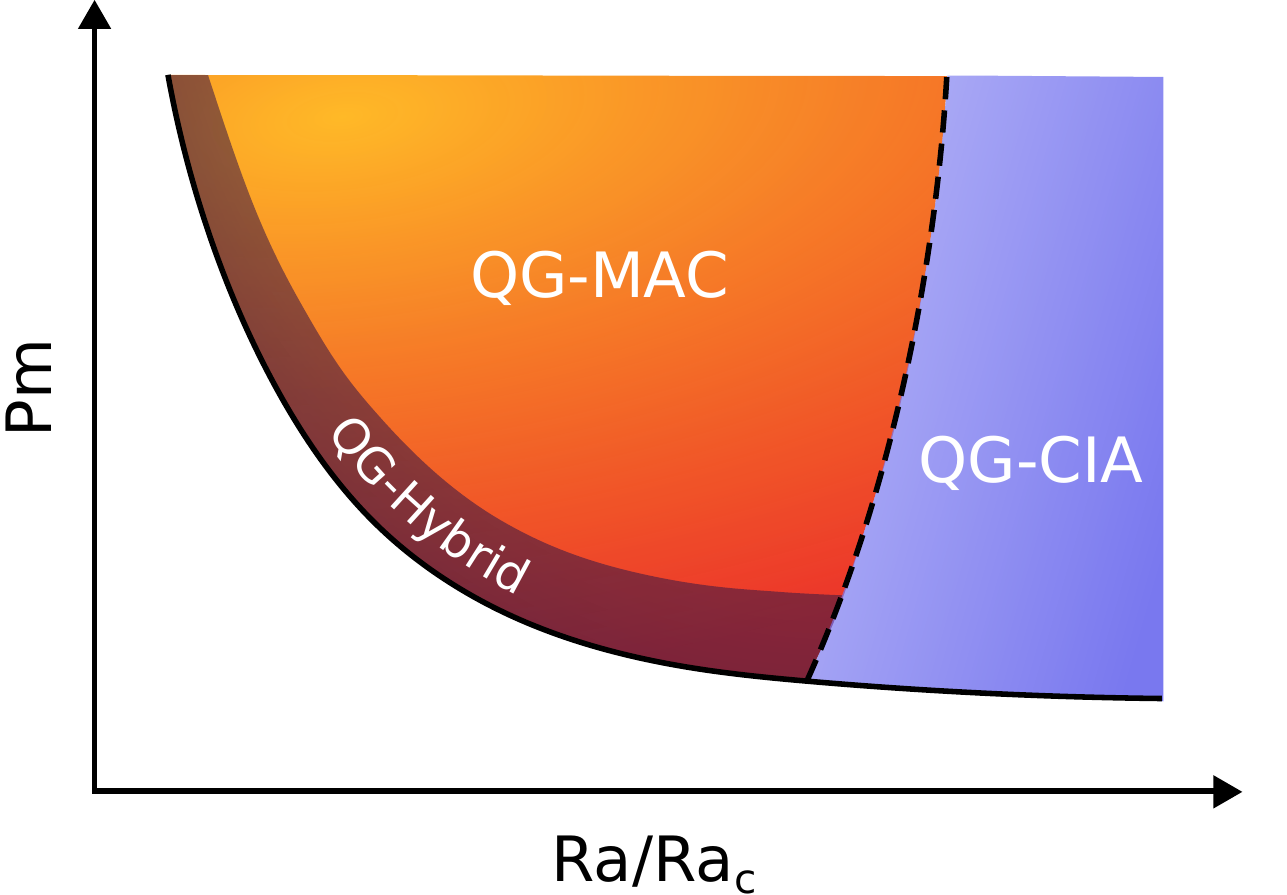}
\caption[]{Sketch of the three different force balance regimes attained in our study independent of the Ekman number. The dashed line marks the transition between the dipolar and multipolar regime.}
\label{discussion_figure}
\end{figure}

By performing a systematic survey of the numerically accessible parameter space, we attempted to provide a better understanding of the force balances controlling the flow dynamics in dynamo models. To this end, we resorted to the length scale based approach introduced by \citet{aubert_etal_2017} and decomposed the amplitudes of the forces into spherical harmonic contributions. We extended this method by additionally looking at the radial dependence of a local force balance measure to analyse possible transitions within the fluid volume. Based on the agreement of the thereby obtained cross-over scales, i.e. scales where three forces are in balance, and the convective scales, we introduced a measure that allows to categorise the force balances into three end-member cases. In agreement with recent studies \citep[e.g.,][]{yadav_etal_2016, aubert_etal_2017, schaeffer_etal_2017}, we find that the majority of the dipole-dominated dynamos are at leading order controlled by a quasi-geostrophic balance between pressure and the Coriolis force. The ageostrophic part of the Coriolis force is then balanced by buoyancy on large scales and the Lorentz force on small scales. This QG-MAC balance seems to be very stable throughout the parameter space, with the exception of boundary regimes. Beyond the transition from the dipolar to the multipolar regime inertial effects become more significant so that inertia now balances the ageostrophic Coriolis force at small scales, while the Lorentz force becomes secondary. The resulting dynamos are therefore governed by a QG-CIA balance at leading order. Close to the onset of dynamo action the Lorentz force also falls onto or below the level of inertia and viscous forces. As a result the QG-MAC balance is lost in these regions for at least parts of the fluid volume. Therefore, one may refer to these dynamos as QG-Hybrid cases. These three different force balance regimes are summarised in a sketch in Fig. \ref{discussion_figure}. Decreasing the Ekman number extends the region of dynamos controlled by a QG-MAC balance towards lower values of $Pm$, while the transition to the QG-CIA regime (the multipolar regime) moves towards higher supercriticalities.

Analysis of the ``strong-fieldness'' and the ratio of magnetic and kinetic energies of the dynamos confirms that the strongest-field dynamos can be found in the parts of the parameter space with high $Pm$ and low $Ra/Ra_{c}$ as suggested by \citet{dormy_2016}. This is mainly a result of the role of inertia being minimised at low supercriticalities. When decreasing the Ekman number, this region of strong-field dynamos will therefore by construction expand to larger supercriticalities due to the increase of rotational effects on the fluid.
These strong-field dynamos do indeed approach magnetostrophy when considering volume-integrated forces. However, the introduction of a finer length scale dependent force balance analysis revealed that these cases are actually controlled by a QG-MAC balance comparable to most dynamo models published to date.

Most dynamo models feature a nearly radially independent force balance. However, in several QG-MAC dynamos a gradual decrease of the convective scale can be observed towards the outer boundary. This can be attributed to a smaller Lorentz force in this region \citep[e.g.,][]{yadav_etal_2016}. Yet, it does not involve a change of the leading-order force balance.
However, by lowering $Pm$ dynamos can be found for which the convective length scale changes very abruptly from elongated structures in the interior of the shell to small-scale convection in the outer part. The 2D analysis of the force balance spectra showed that for these cases the force balance changed from a QG-MAC balance in the interior to a mixed QG-CIA/QG-VAC balance towards the outer boundary. By further decreasing $Pm$ the QG-MAC balance is gradually lost in larger parts of the volume. This stands in contrast to a common strategy in dynamo modelling, which is to decrease $Pm$ as much as possible to approach Earth's core conditions \citep[e.g.,][]{sheyko_etal_2016}. Our results, however, show that following this approach may lead to a significant decrease of the Lorentz force and therefore to the loss of the physically relevant force balance when getting too close to the onset of dynamo action. Strategies such as employed by \citet{dormy_2016} and \citet{aubert_etal_2017}, which aim at preserving the governing force balance when decreasing the Ekman number (by keeping a constant relation between the control parameters), are immune to this.

Future work should focus on asserting the Ekman number dependence of the boundaries of the different force balance regimes. It would also be of interest to perform a more in depth analysis of length scales.

\begin{acknowledgments}
We would like to thank two anonymous reviewers for their constructive comments and Rakesh Yadav for kindly providing the restart files of the dynamo models at $E = 10^{-6}$ that were studied by \citet{yadav_etal_2016}. We acknowledge support from the Fondation Del Duca of Institut de France (JA, 2017 Research Grant). Numerical computations were performed at S-CAPAD, IPGP and using HPC resources from GENCI-CINES (Grants 2016-A0020402122 and 2018-A0020402122). This project has received funding from the European Union's Horizon 2020 research and innovation programme under the Marie Sk\l{}odowska-Curie grant agreement No~665850. This is IPGP contribution 4036.
\end{acknowledgments}

\bibliographystyle{gji}

\label{lastpage}

\appendix
\clearpage
\newpage
\onecolumn
\section{Results of Simulations}
\small
\input{Table.tex}
\twocolumn

\end{document}

%% file: Table.tex
\begin{longtable}{cccccccccccccc}
\caption{Summary of the relevant parameters of the numerical models that were analysed for this study. All dynamo models have been computed with $Pr=1$ and $r_{i}/r_{o}=0.35$. $Nu$ is the Nusselt number, $Rm$ the magnetic Reynolds number, $f_{\mathrm{ohm}}$ the ohmic dissipation fraction and $f_{\mathrm{dip}}$ the relative dipole strength as defined in \citet{christensen_and_aubert_2006}. $\ell_{\perp}$ corresponds to the spherical harmonic degree at which buoyancy and Lorentz forces are of equal magnitude. For cases where this crossing is ill-posed, or the Lorentz force is of the same magnitude or weaker than inertia and/or viscous forces no value is provided.}\\
\toprule
{} &                   $Ra$ &    $Pm$ &   $Nu$ &    $Rm$ & $\Lambda$ & $\mathcal{M}$ & $\delta$ & $\chi_{\mathrm{norm}}$ & $f_{\mathrm{ohm}}$ & $f_{\mathrm{dip}}$ & $\ell_{\mathrm{pol}}$ &    $\ell_{\perp}$ & $(N_{\mathrm{r}},\ell_{\mathrm{max}})$ \\
\midrule
\endfirsthead
\toprule
{} &                   $Ra$ &    $Pm$ &   $Nu$ &    $Rm$ & $\Lambda$ & $\mathcal{M}$ & $\delta$ & $\chi_{\mathrm{norm}}$ & $f_{\mathrm{ohm}}$ & $f_{\mathrm{dip}}$ & $\ell_{\mathrm{pol}}$ &   $\ell_{\perp}$ & $(N_{\mathrm{r}},\ell_{\mathrm{max}})$ \\
\midrule
\endhead
\midrule
\multicolumn{3}{r}{{Continued on next page}} \\
\midrule
\endfoot

\bottomrule
\endlastfoot
\addlinespace[0.1cm]
\multicolumn{14}{c}{$E=10^{-4}$} \\
\addlinespace[0.1cm]
1  &   $1.500\times 10^{6}$ &  12.000 &   1.40 &   209.6 &      8.50 &          23.9 &    20.02 &                   0.07 &               0.60 &               0.82 &                     5 &              5.93 &                               (49, 96) \\
2  &   $1.600\times 10^{6}$ &  12.000 &   1.42 &   205.2 &     11.02 &          32.5 &    22.07 &                   0.09 &               0.66 &               0.84 &                     5 &              5.57 &                               (49, 96) \\
3  &   $1.800\times 10^{6}$ &  12.000 &   1.48 &   203.5 &     19.24 &          57.5 &    28.92 &                   0.03 &               0.73 &               0.82 &                     5 &              4.73 &                               (49, 96) \\
4  &   $2.000\times 10^{6}$ &  12.000 &   1.60 &   238.5 &     25.10 &          54.3 &    30.08 &                   0.05 &               0.73 &               0.77 &                     5 &              4.46 &                              (61, 106) \\
5  &   $2.000\times 10^{6}$ &  15.000 &   1.64 &   312.1 &     35.16 &          55.6 &    29.32 &                   0.04 &               0.71 &               0.72 &                     4 &              3.46 &                              (65, 128) \\
6  &   $2.200\times 10^{6}$ &  12.000 &   1.71 &   267.9 &     29.87 &          51.0 &    29.20 &                   0.03 &               0.72 &               0.76 &                     5 &              4.51 &                              (61, 106) \\
7  &   $2.400\times 10^{6}$ &   5.000 &   1.63 &   153.3 &      3.67 &           8.1 &    13.76 &                   0.12 &               0.52 &               0.86 &                     6 &              8.35 &                               (49, 96) \\
8  &   $2.400\times 10^{6}$ &  12.000 &   1.83 &   299.8 &     34.08 &          46.4 &    26.44 &                   0.06 &               0.71 &               0.74 &                     4 &              4.51 &                              (61, 106) \\
9  &   $2.440\times 10^{6}$ &   2.000 &   1.37 &    66.3 &      0.03 &           0.2 &     0.41 &                   1.03 &               0.04 &               0.97 &                     5 &   --- &                               (41, 85) \\
10 &   $2.750\times 10^{6}$ &  12.000 &   1.99 &   353.7 &     38.94 &          38.0 &    24.12 &                   0.08 &               0.70 &               0.72 &                     5 &              4.53 &                              (61, 106) \\
11 &   $2.790\times 10^{6}$ &   2.000 &   1.55 &    72.3 &      0.42 &           1.6 &     3.47 &                   0.41 &               0.28 &               0.93 &                     6 &             17.32 &                               (41, 85) \\
12 &   $3.200\times 10^{6}$ &   3.000 &   1.78 &   120.5 &      2.12 &           4.4 &    10.20 &                    0.1 &               0.48 &               0.88 &                     7 &              9.87 &                               (41, 96) \\
13 &   $3.200\times 10^{6}$ &   7.000 &   2.07 &   234.8 &     21.26 &          27.5 &    23.86 &                   0.05 &               0.71 &               0.78 &                     6 &              4.94 &                              (61, 106) \\
14 &   $3.200\times 10^{6}$ &  12.000 &   2.18 &   416.0 &     44.11 &          31.2 &    22.06 &                   0.05 &               0.67 &               0.69 &                     5 &              4.61 &                              (61, 106) \\
15 &   $3.500\times 10^{6}$ &   1.000 &   1.86 &    46.4 &      0.45 &           2.1 &     2.64 &                   0.48 &               0.36 &               0.97 &                     8 &             23.87 &                               (41, 85) \\
16 &   $3.750\times 10^{6}$ &   1.000 &   1.97 &    48.4 &      0.69 &           3.0 &     3.72 &                    0.5 &               0.44 &               0.96 &                     7 &             16.54 &                               (41, 85) \\
17 &   $4.830\times 10^{6}$ &   3.000 &   2.46 &   154.8 &      8.54 &          10.9 &    15.17 &                   0.05 &               0.67 &               0.86 &                     6 &              6.70 &                               (41, 96) \\
18 &   $4.830\times 10^{6}$ &   5.000 &   2.65 &   263.1 &     18.52 &          13.6 &    17.97 &                   0.08 &               0.66 &               0.76 &                     7 &              5.20 &                              (61, 106) \\
19 &   $4.830\times 10^{6}$ &   9.500 &   2.80 &   507.7 &     43.33 &          16.2 &    16.74 &                   0.06 &               0.63 &               0.67 &                     5 &              5.00 &                              (49, 106) \\
20 &   $4.880\times 10^{6}$ &   1.000 &   2.35 &    57.8 &      1.55 &           4.7 &     5.57 &                   0.34 &               0.55 &               0.96 &                     7 &             10.92 &                               (41, 85) \\
21 &   $6.500\times 10^{6}$ &   0.500 &   2.84 &    43.6 &      0.49 &           1.3 &     1.23 &                   1.29 &               0.31 &               0.97 &                     5 &             --- &                               (41, 85) \\
22 &   $7.500\times 10^{6}$ &   0.500 &   3.25 &    50.4 &      0.60 &           1.2 &     1.22 &                   1.01 &               0.31 &               0.96 &                     6 &   --- &                               (41, 64) \\
23 &   $7.500\times 10^{6}$ &   1.000 &   3.16 &    84.6 &      2.70 &           3.8 &     5.01 &                   0.08 &               0.54 &               0.95 &                     6 &             10.91 &                               (41, 85) \\
24 &   $7.500\times 10^{6}$ &   3.000 &   3.44 &   249.7 &     11.87 &           5.8 &    10.31 &                    0.1 &               0.59 &               0.78 &                     7 &              8.19 &                              (49, 106) \\
25 &   $7.500\times 10^{6}$ &   9.000 &   3.64 &   726.6 &     53.23 &           9.2 &    12.92 &                   0.12 &               0.56 &               0.63 &                     6 &              5.67 &                              (49, 106) \\
26 &   $7.500\times 10^{6}$ &  12.000 &   3.67 &   957.0 &     77.88 &          10.3 &    14.16 &                   0.05 &               0.55 &               0.60 &                     6 &              5.30 &                              (65, 128) \\
27 &   $7.500\times 10^{6}$ &  15.000 &   3.68 &  1195.4 &     99.66 &          10.6 &    13.38 &                   0.08 &               0.53 &               0.58 &                     7 &              5.08 &                              (65, 133) \\
28 &   $8.250\times 10^{6}$ &   7.000 &   3.83 &   622.6 &     39.97 &           7.3 &    11.94 &                   0.09 &               0.56 &               0.64 &                     6 &              6.32 &                              (49, 106) \\
29 &   $8.500\times 10^{6}$ &   0.500 &   3.64 &    57.9 &      0.64 &           1.0 &     1.08 &                   1.01 &               0.29 &               0.95 &                     7 &             --- &                               (41, 85) \\
30 &   $1.125\times 10^{7}$ &   1.000 &   4.28 &   127.0 &      3.41 &           2.1 &     3.58 &                   0.08 &               0.48 &               0.91 &                     8 &             12.29 &                               (41, 85) \\
31 &   $1.125\times 10^{7}$ &   3.000 &   4.51 &   366.1 &     14.37 &           3.2 &     6.95 &                   0.15 &               0.53 &               0.73 &                     6 &              8.92 &                              (49, 106) \\
32 &   $1.125\times 10^{7}$ &   5.000 &   4.59 &   594.4 &     29.30 &           4.2 &     8.89 &                   0.18 &               0.53 &               0.66 &                     7 &              8.40 &                              (49, 106) \\
33 &   $1.125\times 10^{7}$ &   7.000 &   4.61 &   814.6 &     46.53 &           5.0 &    10.36 &                   0.15 &               0.53 &               0.62 &                     7 &              6.78 &                              (61, 128) \\
34 &   $1.500\times 10^{7}$ &   1.000 &   5.90 &   227.7 &      0.49 &           0.1 &     0.36 &                   5.48 &               0.11 &               0.31 &                     6 &   --- &                               (41, 85) \\
35 &   $1.500\times 10^{7}$ &   2.000 &   5.41 &   324.0 &      9.27 &           1.8 &     4.12 &                   0.16 &               0.47 &               0.77 &                    12 &             11.23 &                               (41, 85) \\
36 &   $1.500\times 10^{7}$ &   3.000 &   5.46 &   473.0 &     16.21 &           2.2 &     5.45 &                   0.15 &               0.48 &               0.71 &                     9 &             10.38 &                              (49, 106) \\
37 &   $1.500\times 10^{7}$ &   5.000 &   5.46 &   757.2 &     33.83 &           3.0 &     6.87 &                   0.19 &               0.48 &               0.64 &                     6 &              8.82 &                              (49, 106) \\
38 &   $1.500\times 10^{7}$ &   7.000 &   5.47 &  1038.1 &     53.11 &           3.5 &     7.77 &                   0.16 &               0.48 &               0.59 &                     8 &              8.35 &                              (61, 128) \\
39 &   $1.500\times 10^{7}$ &   9.000 &   5.45 &  1309.4 &     74.79 &           4.0 &     8.80 &                   0.14 &               0.47 &               0.57 &                     7 &              8.28 &                              (65, 133) \\
40 &   $1.500\times 10^{7}$ &  15.000 &   5.38 &  2108.2 &    147.91 &           5.1 &     9.75 &                   0.12 &               0.42 &               0.52 &                     6 &              6.71 &                              (65, 133) \\
41 &   $1.750\times 10^{7}$ &   1.000 &   6.53 &   255.2 &      0.66 &           0.1 &     0.38 &                   6.08 &               0.12 &               0.31 &                     6 &   --- &                              (61, 106) \\
42 &   $1.750\times 10^{7}$ &   2.000 &   6.09 &   381.5 &      8.83 &           1.2 &     3.28 &                   0.11 &               0.43 &               0.75 &                    14 &             13.01 &                              (61, 106) \\
43 &   $1.750\times 10^{7}$ &   4.560 &   5.98 &   785.2 &     32.62 &           2.4 &     6.14 &                   0.19 &               0.47 &               0.65 &                    11 &              8.91 &                              (61, 106) \\
44 &   $1.750\times 10^{7}$ &   8.000 &   5.96 &  1327.8 &     69.66 &           3.2 &     7.85 &                   0.15 &               0.45 &               0.57 &                     8 &              8.43 &                              (61, 128) \\
45 &   $2.100\times 10^{7}$ &   2.000 &   7.21 &   522.6 &      4.05 &           0.3 &     1.21 &                   1.61 &               0.26 &               0.28 &                     6 &   --- &                              (61, 128) \\
46 &   $2.100\times 10^{7}$ &   7.000 &   6.61 &  1353.1 &     63.08 &           2.4 &     6.50 &                   0.17 &               0.45 &               0.58 &                     6 &              8.69 &                              (81, 133) \\
\addlinespace[0.1cm]
\multicolumn{14}{c}{$E=10^{-5}$} \\
\addlinespace[0.1cm]
47 &   $3.300\times 10^{7}$ &   5.000 &   1.83 &   271.5 &      3.74 &          27.4 &    36.83 &                   0.03 &               0.75 &               0.27 &                     8 &              7.70 &                              (81, 133) \\
48 &   $3.300\times 10^{7}$ &   7.000 &   1.87 &   346.2 &      7.15 &          42.5 &    37.42 &                   0.05 &               0.78 &               0.75 &                     9 &              6.79 &                              (97, 133) \\
49 &   $4.000\times 10^{7}$ &   2.000 &   1.33 &   104.1 &      0.09 &           1.7 &     3.54 &                   0.67 &               0.26 &               0.97 &                    12 &             31.88 &                              (81, 133) \\
50 &   $4.000\times 10^{7}$ &   7.000 &   2.12 &   303.1 &     23.01 &         180.8 &    61.43 &                   0.02 &               0.88 &               0.84 &                     6 &              4.92 &                              (97, 133) \\
51 &   $5.000\times 10^{7}$ &   1.000 &   1.52 &    72.6 &      0.09 &           1.8 &     3.03 &                   0.58 &               0.33 &               0.96 &                    14 &             39.73 &                              (81, 133) \\
52 &   $5.000\times 10^{7}$ &   2.000 &   1.43 &   132.9 &      0.16 &           1.8 &     5.10 &                   0.68 &               0.30 &               0.87 &                    12 &             28.63 &                              (81, 133) \\
53 &   $5.000\times 10^{7}$ &   3.000 &   2.05 &   240.9 &      2.21 &          12.1 &    23.33 &                   0.06 &               0.69 &               0.76 &                     8 &             10.46 &                              (81, 133) \\
54 &   $6.000\times 10^{7}$ &   5.000 &   2.81 &   337.2 &     26.37 &         117.5 &    57.82 &                   0.01 &               0.87 &               0.80 &                     7 &              4.93 &                              (97, 170) \\
55 &   $7.500\times 10^{7}$ &   0.500 &   2.22 &    60.8 &      0.20 &           2.8 &     2.96 &                   0.98 &               0.44 &               0.90 &                    13 &               --- &                              (81, 133) \\
56 &   $7.500\times 10^{7}$ &   2.000 &   2.49 &   194.1 &      3.27 &          17.6 &    33.57 &                   0.05 &               0.77 &               0.91 &                     8 &              8.96 &                              (81, 133) \\
57 &   $8.000\times 10^{7}$ &   1.000 &   2.46 &   120.0 &      0.93 &           6.5 &    12.55 &                   0.13 &               0.63 &               0.84 &                    10 &             12.50 &                              (81, 133) \\
58 &   $1.000\times 10^{8}$ &   1.000 &   3.06 &   140.5 &      1.76 &           9.1 &    14.19 &                   0.11 &               0.69 &               0.91 &                     9 &             11.10 &                              (81, 133) \\
59 &   $1.000\times 10^{8}$ &   2.000 &   3.67 &   225.5 &     11.35 &          45.3 &    36.12 &                   0.09 &               0.86 &               0.91 &                     6 &              6.39 &                              (81, 133) \\
60 &   $1.000\times 10^{8}$ &   3.000 &   3.90 &   341.1 &     21.15 &          55.4 &    37.40 &                   0.04 &               0.85 &               0.83 &                     6 &              5.65 &                              (97, 170) \\
61 &   $1.000\times 10^{8}$ &   7.000 &   4.07 &   779.9 &     63.63 &          74.9 &    31.72 &                   0.04 &               0.80 &               0.75 &                     6 &              4.96 &                              (97, 170) \\
62 &   $1.100\times 10^{8}$ &   0.200 &   3.01 &    39.6 &      0.07 &           0.9 &     0.45 &                    --- &               0.23 &               0.99 &                    12 &  --- &                              (97, 170) \\
63 &   $1.500\times 10^{8}$ &   0.200 &   4.10 &    51.0 &      0.19 &           1.4 &     0.86 &                   1.41 &               0.33 &               0.96 &                    14 &  --- &                              (97, 170) \\
64 &   $1.500\times 10^{8}$ &   0.300 &   4.28 &    72.6 &      0.42 &           2.4 &     2.11 &                   0.31 &               0.44 &               0.91 &                    13 &               --- &                              (81, 133) \\
65 &   $1.500\times 10^{8}$ &   0.500 &   4.45 &   116.9 &      1.03 &           3.8 &     4.54 &                   0.09 &               0.56 &               0.90 &                    12 &             16.08 &                              (81, 133) \\
66 &   $1.500\times 10^{8}$ &   2.000 &   5.13 &   346.3 &     15.87 &          26.8 &    29.72 &                   0.04 &               0.82 &               0.87 &                     7 &              7.39 &                              (97, 170) \\
67 &   $2.000\times 10^{8}$ &   0.150 &   6.31 &    55.8 &      0.24 &           1.2 &     0.62 &                   3.78 &               0.34 &               0.98 &                    12 &   --- &                              (97, 170) \\
68 &   $2.000\times 10^{8}$ &   1.000 &   6.02 &   240.4 &      7.03 &          12.3 &    15.42 &                   0.06 &               0.77 &               0.94 &                     8 &              9.60 &                              (97, 170) \\
69 &   $2.000\times 10^{8}$ &   2.000 &   6.31 &   451.4 &     19.88 &          19.7 &    24.82 &                   0.05 &               0.80 &               0.84 &                     9 &              8.18 &                              (97, 170) \\
70 &   $2.000\times 10^{8}$ &   5.000 &   6.57 &  1107.2 &     63.20 &          26.1 &    24.92 &                   0.05 &               0.74 &               0.74 &                     7 &              6.60 &                              (97, 170) \\
71 &   $3.000\times 10^{8}$ &   0.150 &   9.90 &    78.8 &      0.55 &           1.3 &     0.69 &                   2.32 &               0.43 &               0.96 &                    13 &   --- &                              (97, 170) \\
72 &   $3.000\times 10^{8}$ &   0.200 &   9.89 &    99.3 &      0.99 &           2.0 &     1.33 &                   0.35 &               0.51 &               0.96 &                    14 &               --- &                              (97, 170) \\
73 &   $3.000\times 10^{8}$ &   0.300 &   9.34 &   136.8 &      1.70 &           2.8 &     2.48 &                    0.1 &               0.58 &               0.94 &                    12 &             15.72 &                              (97, 170) \\
74 &   $3.000\times 10^{8}$ &   0.500 &   8.43 &   202.0 &      3.16 &           3.9 &     5.11 &                   0.08 &               0.65 &               0.95 &                    12 &             12.83 &                              (97, 192) \\
75 &   $4.000\times 10^{8}$ &   0.100 &  12.38 &    89.3 &      0.10 &           0.1 &     0.22 &                    --- &               0.19 &               0.25 &                     7 &   --- &                              (97, 192) \\
76 &   $4.000\times 10^{8}$ &   0.150 &  12.52 &    98.8 &      0.89 &           1.4 &     0.84 &                   1.27 &               0.48 &               0.97 &                    12 &   --- &                              (97, 192) \\
77 &   $4.000\times 10^{8}$ &   1.000 &  10.33 &   432.5 &     13.02 &           7.0 &     9.80 &                   0.07 &               0.74 &               0.90 &                    11 &              9.71 &                             (121, 256) \\
78 &   $4.000\times 10^{8}$ &   2.000 &  10.55 &   845.0 &     30.59 &           8.6 &    13.97 &                   0.07 &               0.73 &               0.81 &                    10 &              8.96 &                             (121, 256) \\
79 &   $5.000\times 10^{8}$ &   0.070 &  14.70 &    77.9 &      0.07 &           0.1 &     0.16 &                    --- &               0.17 &               0.24 &                     7 &    --- &                              (97, 192) \\
80 &   $5.000\times 10^{8}$ &   0.150 &  14.28 &   116.8 &      1.07 &           1.2 &     0.87 &                   0.91 &               0.49 &               0.97 &                    12 &   --- &                              (97, 192) \\
81 &   $7.000\times 10^{8}$ &   0.200 &  18.93 &   236.2 &      0.86 &           0.3 &     0.59 &                   9.28 &               0.36 &               0.30 &                     9 &   --- &                             (121, 256) \\
82 &   $7.000\times 10^{8}$ &   0.500 &  15.35 &   375.5 &      8.28 &           2.9 &     4.70 &                   0.15 &               0.68 &               0.94 &                    11 &              8.95 &                             (121, 256) \\
83 &   $7.000\times 10^{8}$ &   1.440 &  15.64 &  1018.0 &     27.55 &           3.8 &     6.94 &                   0.11 &               0.68 &               0.85 &                    12 &             10.42 &                             (121, 256) \\
84 &   $7.000\times 10^{8}$ &   5.000 &  15.18 &  3286.8 &    131.81 &           6.2 &     7.86 &                   0.12 &               0.56 &               0.70 &                     9 &              8.87 &                             (121, 256) \\
\addlinespace[0.1cm]
\multicolumn{14}{c}{$E=10^{-6}$} \\
\addlinespace[0.1cm]
85 &   $8.000\times 10^{8}$ &   2.000 &   1.37 &   243.0 &      0.04 &           1.5 &     3.27 &                   0.68 &               0.23 &               0.91 &                    39 &             72.58 &                             (161, 341) \\
86 &   $1.000\times 10^{9}$ &   1.000 &   1.67 &   141.9 &      0.15 &           7.6 &    15.12 &                   0.76 &               0.63 &               0.81 &                    29 &             25.77 &                             (161, 426) \\
87 &   $2.000\times 10^{9}$ &   0.150 &   2.84 &    57.1 &      0.04 &           1.8 &     0.82 &                    --- &               0.32 &               0.93 &                    22 &  --- &                             (161, 341) \\
88 &   $2.000\times 10^{9}$ &   0.250 &   3.42 &    97.9 &      0.14 &           3.5 &     3.70 &                   1.33 &               0.49 &               0.77 &                    24 &               --- &                             (193, 426) \\
89 &   $2.000\times 10^{9}$ &   0.500 &   4.93 &   176.5 &      1.22 &          19.8 &    19.65 &                   0.21 &               0.82 &               0.90 &                    17 &             16.90 &                             (161, 426) \\
90 &   $2.000\times 10^{9}$ &   1.000 &   4.98 &   340.9 &      2.65 &          22.9 &    37.09 &                   0.03 &               0.84 &               0.92 &                    12 &             13.22 &                             (193, 426) \\
91 &   $2.000\times 10^{9}$ &   2.000 &   5.69 &   525.9 &     11.74 &          85.3 &    83.90 &                   0.02 &               0.91 &               0.93 &                    11 &              8.86 &                             (193, 426) \\
92 &   $2.800\times 10^{9}$ &   0.500 &   6.98 &   259.5 &      1.72 &          12.8 &    16.58 &                   0.03 &               0.78 &               0.88 &                    18 &             15.63 &                             (193, 426) \\
93 &   $5.500\times 10^{9}$ &   0.400 &  12.78 &   358.0 &      3.61 &          11.3 &    12.36 &                   0.12 &               0.80 &               0.88 &                    22 &             11.07 &                             (289, 426) \\
94 &  $1.000\times 10^{10}$ &   0.500 &  20.05 &   580.9 &     11.53 &          17.2 &    17.52 &                   0.02 &               0.87 &               0.94 &                    13 &              9.64 &                             (321, 512) \\
95 &  $2.660\times 10^{10}$ &   0.456 &  41.22 &  1085.6 &     30.10 &          11.8 &    11.11 &                   0.04 &               0.90 &               0.94 &                    11 &              8.76 &                             (361, 512) \\
\end{longtable}